\documentclass[12pt]{iopart}
\usepackage{graphicx} 

\begin{document}

\title{Dynamics of decoherence without dissipation in a 
squeezed thermal bath}

\author{Subhashish Banerjee}
\address{Raman Research Institute, Bangalore - 560080, 
India} 
\author{R Ghosh \cite{rg}} 
\address{School of Physical Sciences, Jawaharlal Nehru 
University, New Delhi - 110067, India} 

\date{17 September 2007}

\begin{abstract}
We study a generic open  quantum system where the coupling between the
system  and  its  environment   is  of  an  energy-preserving  quantum
nondemolition (QND)  type. We obtain  the general master  equation for
the  evolution of  such a  system under  the influence  of  a squeezed
thermal bath of harmonic oscillators.  From the master equation it can
be seen explicitly that  the process involves decoherence or dephasing
without any dissipation of energy. We work out the decoherence-causing
term in the high and zero temperature limits and check that they match
with known  results for the case of  a thermal bath. The  decay of the
coherence is quantified as well  by the dynamics of the linear entropy
of  the  system under  various  environmental  conditions.  We make  a
comparison  of  the quantum  statistical  properties  between QND  and
dissipative types of evolution using  a system of two-level atom and a
harmonic oscillator.
\end{abstract} 

\submitto{\JPA}


\maketitle

\section{Introduction}

The concept of `open' quantum systems  is a ubiquitous one in that all
real  systems of  interest are  open systems,  each surrounded  by its
environment,  which  affects   its  dynamics.   Caldeira  and  Leggett
\cite{cl83}  used  the  influence  functional  approach  developed  by
Feynman and Vernon \cite{fv63}  to discuss quantum dissipation via the
paradigm  of  quantum  Brownian  motion  (QBM) of  a  simple  harmonic
oscillator in a harmonic  oscillator environment. The influence of the
environment on  the reduced dynamics  of the system was  quantified by
the influence  functional. Dissipation  of the system  originates from
the  transfer of energy  from the  system of  interest to  the `large'
environment. The  energy, once transferred,  is not given back  to the
system within any time of physical relevance. In the original model of
the QBM,  the system  and its environment  were taken to  be initially
uncorrelated. The treatment was  extended to the physically reasonable
initial condition of  a mixed state of the  system and its environment
by Hakim  and Ambegaokar \cite{ha85}, Smith  and Caldeira \cite{sc87},
Grabert,  Schramm  and Ingold  \cite{gsi88},  and  Banerjee and  Ghosh
\cite{bg03} among  others. Haake and Reibold \cite{hr85},  and Hu, Paz
and  Zhang \cite{hpz92}  obtained  an exact  master  equation for  the
quantum  Brownian  particle for  a  general  spectral  density of  the
environment.

The spectacular progress in manipulation of quantum states of 
matter and applications in quantum information processing have 
resulted in a renewed demand for understanding and control of 
the environmental impact in such open quantum systems. For such 
systems, there exists an important class of energy-preserving 
measurements in which dephasing occurs without damping of the 
system. This may be achieved with a particular type of coupling 
between the system and its environment, viz., when the 
Hamiltonian $H_S$ of the system commutes with the Hamiltonian 
$H_{SR}$ describing the system-reservoir interaction, i.e., 
$H_{SR}$ is a constant of motion generated by $H_S$ 
\cite{sgc96,mp98,gkd01}. This condition describes a 
particular type of quantum nondemolition (QND) measurement 
scheme. 

In general, a class of observables that may be measured 
repeatedly with arbitrary precision, with the influence of the 
measurement apparatus on the system being confined strictly to 
the conjugate observables, is called QND or back-action evasive 
observables \cite{bvt80,bk92,wm94,zu84,mk00}. Such a 
measurement scheme was originally introduced in the context of 
the detection of gravitational waves \cite{ct80,bo96}. The 
dynamics of decoherence in continuous QND measurements applied 
to a system of two-level atom interacting with a stationary 
quantized electromagnetic field through a dispersive coupling 
has been studied by Onofrio and Viola \cite{vo98}. In addition 
to its relevance in ultrasensitive measurements, a QND scheme 
provides a way to prepare quantum mechanical states which may 
otherwise be difficult to create, such as Fock states with a 
specific number of particles. It has been shown that the 
accuracy of atomic interferometry can be improved by using QND 
measurements of the atomic populations at the inputs to the 
interferometer \cite{kbm98}. QND systems have also been 
proposed for engineering quantum dynamical evolution of a 
system with the help of a quantum meter \cite{ca05}. We have 
recently studied such QND open system Hamiltonians for two 
different models of the environment describable as baths of 
either oscillators or spins, and found an interesting 
connection between the energy-preserving QND Hamiltonians and 
the phase space area-preserving canonical transformations 
\cite{sb07}. 

In this paper we wish to study the dynamics of decoherence in a 
generic open quantum system where the coupling between the 
system and its environment is of an energy-preserving QND type. 
The bath is taken to be initially in a squeezed thermal state, 
from which the common thermal bath results may be easily 
extracted by setting the squeezing parameters to zero. When the 
quantum fluctuations of the heat bath are squeezed, it has been 
shown by Kennedy and Walls \cite{kw88} that the macroscopic 
superposition of states of light is preserved in the presence 
of dissipation. These authors have shown that the squeezed bath 
is more efficient than the thermal bath for optical 
quadrature-phase quantum measurements, and may also be used to 
prepare the states with low quantum noise in one quadrature 
phase, at least in the high-frequency regime. The advantage of 
using a squeezed thermal bath over an ordinary 
phase-insensitive thermal bath is that the decay rate of 
quantum coherences can be suppressed in a squeezed bath leading 
to preservation of nonclassical effects \cite{kb93}. Such a 
bath has also been shown to modify the evolution of the 
geometric phase of two-level atomic systems \cite{br06}. In our 
present problem, we wish to systematically probe the effect of 
phase-sensitivity of the bath, and quantify the pattern of 
progressive decay of coherence of the system, both at high 
temperatures as well as arbitrary low temperatures, when a 
quantum nondemolition coupling is adopted. We wish to hence 
compare and contrast the quantum statistical mechanical 
features (viz. the nature of the noise channels) of the QND 
type of evolution with that of the dissipative evolution 
\cite{hpz92,hm94,bk05,b04,bp02} of the general Lindblad 
form for a two-level system or the specific QBM form for a 
harmonic oscillator. 

The plan of the paper is as follows. In section 2 we obtain 
the master equation for a generic system interacting with its 
environment by a QND type of coupling. For simplicity, we take 
the system and its environment to be initially separable. In 
section 2.1, the master equation is obtained for the case of a 
bosonic bath of harmonic oscillators. For the sake of 
completeness, we briefly compare the results for the oscillator 
bath with that for a bath of two-level systems in section 2.2. 
In section 3, we analyze the dynamics of decoherence, first 
by looking at the term causing decoherence in the system master 
equation for the bosonic bath of harmonic oscillators obtained 
in section 2.1, and explicitly solve it for the 
high-temperature and the zero-temperature cases. We then set up 
a quantitative `measure of coherence' related to the linear 
entropy $S(t)$, the dynamics of which is also studied for the 
zero- as well as the high-temperature cases for different 
degrees of squeezing of the bath. In section 4, the quantum 
statistical mechanical properties underlying the QND and 
dissipative processes are studied on a general footing for a 
two-level atomic system (section 4.1) and a harmonic oscillator 
system (section 4.2). For the two-level system, the dissipative 
process is taken to be that generated by a standard Lindblad 
equation while for the harmonic oscillator system, the model 
studied is that of QBM. In section 5 we present our conclusions. 

\section{Master equation}

Here we present the master equation for a system interacting 
with its environment by a coupling of the energy-preserving QND 
type where the environment is a bosonic bath of harmonic 
oscillators initially in a squeezed thermal state, decoupled 
from the system. We also take up the case where the environment 
is composed of a bath of two-level systems, and compare the two 
cases. 

\subsection{Bath of harmonic oscillators}

We consider the Hamiltonian
\begin{eqnarray}
H & = & H_S + H_R + H_{SR} \nonumber \\ & = & H_S + 
\sum\limits_k \hbar \omega_k b^{\dagger}_k b_k + H_S 
\sum\limits_k g_k (b_k+b^{\dagger}_k) + H^2_S \sum\limits_k 
{g^2_k \over \hbar \omega_k}. 
\end{eqnarray} 
Here $H_S$, $H_R$ and $H_{SR}$ stand for the Hamiltonians of 
the system, reservoir and system-reservoir interaction, 
respectively. $H_S$ is a generic system Hamiltonian which can 
be specified depending on the physical situation. 
$b^{\dagger}_k$, $b_k$ denote the creation and annihilation 
operators for the reservoir oscillator of frequency $\omega_k$, 
$g_k$ stands for the coupling constant (assumed real) for the 
interaction of the oscillator field with the system. The last 
term on the right-hand side of Eq. (1) is a renormalization 
inducing `counter term'. Since $[H_S, H_{SR}]=0$, the 
Hamiltonian (1) is of QND type. The system-plus-reservoir 
composite is closed and hence obeys a unitary evolution given 
by 
\begin{equation}
\rho (t) = e^{- iHt / \hbar} \rho (0) e^{iHt / \hbar} , 
\end{equation}
where
\begin{equation}
\rho (0) = \rho^s (0) \rho_R (0),
\end{equation}
i.e., we assume separable initial conditions. In order to 
obtain the reduced dynamics of the system alone, we trace over 
the reservoir variables. The matrix elements of the reduced 
density matrix in the system eigenbasis are 
\begin{eqnarray}
\rho^s_{nm}(t) & = & e^{- i (E_n-E_m)t /\hbar}~~ e^{- i 
(E^2_n-E^2_m)/\hbar \sum\limits_k (g^2_k t / \hbar \omega_k)} 
\nonumber \\ & & \times \rm{Tr}_R \left[ e^{- i H_n t/\hbar } 
\rho_R (0) e^{i H_m t/\hbar } \right] \rho^s_{nm}(0) , \label{rhot}
\end{eqnarray}
where $E_n$'s are the eigenvalues of the system Hamiltonian. In 
Eq. (\ref{rhot}), $\rho_R (0)$ is the initial density matrix of the
reservoir which we take to be a squeezed thermal bath given by
\begin{equation}
\rho_R(0) = S(r, \Phi) \rho_{th} S^{\dagger} (r, \Phi), \label{rhorin}
\end{equation}
where
\begin{equation}
\rho_{th} = \prod_k \left[ 1 - e^{- \beta \hbar \omega_k} 
\right] e^{-\beta \hbar \omega_k b^{\dagger}_k b_k} \label{rhoth}
\end{equation}
is the density matrix of the thermal bath at temperature $T$, 
with $\beta \equiv 1/(k_B T)$, $k_B$ being the Boltzmann 
constant, and 
\begin{equation}
S(r_k, \Phi_k) = \exp \left[ r_k \left( {b^2_k \over 2} e^{-2i 
\Phi_k} - {b^{\dagger 2}_k \over 2} e^{2i \Phi_k} \right) 
\right] \label{sqop}
\end{equation}
is the squeezing operator with $r_k$, $\Phi_k$ being the 
squeezing parameters \cite{cs85}. In Eq. (\ref{rhot}), 
\begin{equation}
H_n = \sum\limits_k \left[ \hbar \omega_k b^{\dagger}_k b_k + 
E_n g_k (b_k + b^{\dagger}_k) \right]. \label{hn}
\end{equation}
Following the steps of the derviation as shown in \ref{sec:deriv}, 
the reduced density matrix (4) of the system is obtained as
\begin{eqnarray}
\fl \rho^s_{nm} (t) & = & e^{- i (E_n - E_m)t /\hbar} e^{-
i(E^2_n - E^2_m) \sum\limits_k (g^2_k \sin (\omega_kt)/\hbar^2 
\omega^2_k)} \nonumber \\ & & \times \exp \Bigg[ - {1 
\over 2} (E_m - E_n)^2 \sum\limits_k {g^2_k \over \hbar^2 
\omega^2_k} \coth \left( {\beta \hbar \omega_k \over 2} \right) 
\nonumber \\ & & \times \left| (e^{i\omega_k t} - 1) \cosh 
(r_k) + (e^{-i\omega_kt} - 1) \sinh (r_k) e^{2i \Phi_k} 
\right|^2 \Bigg] \rho^s_{nm} (0). \label{h1} 
\end{eqnarray}

Differentiating Eq. (\ref{h1}) with respect to time we obtain 
the master equation giving the system evolution under the 
influence of the environment as 
\begin{equation}
\fl \dot{\rho}^s_{nm} (t) = \left[ -{i \over \hbar} (E_n - E_m) + i 
\dot{\eta} (t) (E^2_n - E^2_m) - (E_n - E_m)^2 \dot{\gamma} (t) 
\right] \rho^s_{nm} (t), \label{h2} 
\end{equation}
where
\begin{equation}
\eta (t) = - \sum\limits_k {g^2_k \over \hbar^2 \omega^2_k} 
\sin (\omega_k t), \label{h3} 
\end{equation}
and
\begin{equation}
\fl \gamma (t) = {1 \over 2} \sum\limits_k {g^2_k \over \hbar^2 
\omega^2_k} \coth \left( {\beta \hbar \omega_k \over 2} \right) 
\left| (e^{i\omega_k t} - 1) \cosh (r_k) + (e^{-i\omega_k t} - 
1) \sinh (r_k) e^{2i \Phi_k} \right|^2. \label{h4} 
\end{equation}

For the case of zero squeezing, $r = \Phi = 0$, and $\gamma 
(t)$ given by Eq. (\ref{h4}) reduces to the expression obtained 
earlier \cite{sgc96,mp98,gkd01} for the case of a thermal 
bath. It can be seen that $\eta (t)$ (\ref{h3}) is independent 
of the bath initial conditions and hence remains the same as 
for the thermal bath. Comparing the master equation obtained 
for the case of a QND coupling to the bath (\ref{h2}) with the 
master equation obtained in the case of QBM as in 
Refs. \cite{hm94,bk05,b04}, where the master equation was 
obtained for the QBM of the system of a harmonic oscillator in 
a squeezed thermal bath, we find that the term responsible for 
decoherence in the QND case is given by $\dot{\gamma}(t)$. It 
is interesting to note that in contrast to the QBM case, here 
there is no term governing dissipation. Also missing are the 
various other diffusion terms, viz. those responsible for 
promoting diffusion in $p^2$ and those responsible for 
diffusion in $xp + px$, the so-called anomalous diffusion 
terms. Also note that in the exponent of the third exponential 
on the right-hand side of Eq. (\ref{h1}), responsible for the 
decay of coherences, the coefficient of $\gamma (t)$ is 
dependent on the eigenvalues $E_n$ of the `conserved pointer 
observable' operator which in this case is the system 
Hamiltonian itself. This reiterates the observation that the 
decay of coherence in a system interacting with its bath via a 
QND interaction depends on the conserved pointer observable and 
the bath coupling parameters \cite{mp98}. 

\subsection{Bath of two-level systems}
 
We briefly take up the case of a bath of two-level systems to 
illustrate in a transparent manner its difference with a bath 
of harmonic oscillators. The Hamiltonian considered is 
\begin{eqnarray}
H & = & H_S + H_R + H_{SR} \nonumber \\ & = & H_S + 
\sum\limits_k \omega_k \sigma_{zk} + H_S \sum\limits_k C_k 
\sigma_{xk}. \label{twolv1}
\end{eqnarray}
Since $[H_S, H_{SR}]=0$, the Hamiltonian (\ref{twolv1}) is of a QND
type. Starting from the unitary evolution of the entire closed system
and then tracing over the bath variables, we obtain the reduced
density matrix in the system eigenbasis as
\begin{equation}
\rho^s_{nm} (t) = e^{- i (E_n - E_m)t/\hbar} \rm{Tr}_R \left[ 
e^{i H_m t/\hbar} e^{- i H_n t/\hbar} \rho_R(0) 
\right] \rho^s_{nm}(0) , \label{twolv2}
\end{equation}
where
\begin{equation}
H_n = \sum\limits_k \left[ \omega_k \sigma_{zk} + E_n C_k 
\sigma_{xk} \right] = \sum\limits_k O_k (E_n). \label{twolv3}
\end{equation}
Using the properties of the $\sigma_z, \sigma_x$ matrices it 
can be seen that 
\begin{equation}
e^{i O_k(E_m)t} = \cos \left( \omega'_k (E_m)t \right) + {i 
\sin \left( \omega'_k (E_m)t \right) \over \omega'_k (E_m)} 
(\omega_k \sigma_{zk} + E_m C_k \sigma_{xk}) , \label{twolv4}
\end{equation}
where
\begin{equation}
\omega'_k (E_m) = \sqrt{\omega^2_k + E^2_m C^2_k}. \label{twolv5}
\end{equation}
Thus
\begin{eqnarray}
\fl e^{i O_k(E_m)t} e^{-i O_k(E_n)t} & = & \cos \left( \omega'_k 
(E_m)t \right) \cos \left( \omega'_k (E_n)t \right) \nonumber 
\\ & & + {\sin \left(\omega'_k (E_m)t\right) \sin \left( 
\omega'_k (E_n)t\right) \over \omega'_k (E_m) \omega'_k (E_n)} 
\left( \omega^2_k + E_m E_n C^2_k \right) \nonumber \\ & & + 
{\sin \left( \omega'_k (E_m)t\right) \sin \left( \omega'_k 
(E_n)t \right) \over \omega'_k (E_m) \omega'_k (E_n)} \omega_k 
C_k (E_n - E_m) \sigma_{zk} \sigma_{xk} \nonumber \\ & & - {i 
\cos \left( \omega'_k (E_m)t\right) \sin \left( \omega'_k 
(E_n)t\right) \over \omega'_k (E_n)} (\omega_k \sigma_{zk} + 
E_n C_k \sigma_{xk}) \nonumber \\ & & + {i \cos \left( 
\omega'_k (E_n)t\right) \sin \left( \omega'_k (E_m)t\right) 
\over \omega'_k (E_m)} (\omega_k \sigma_{zk} + E_m C_k 
\sigma_{xk}). \label{twolv6}
\end{eqnarray}
Using (\ref{twolv6}) in (\ref{twolv2}), it can be seen that only the
first two terms on the right-hand side of Eq. (\ref{twolv6}) contribute and the reduced density matrix of the system becomes
\begin{eqnarray}
\rho^s_{nm}(t) & = & e^{- i (E_n - E_m)t/\hbar} \prod_k 
\Bigg[ \cos \left( \omega'_k (E_m)t\right) \cos \left( 
\omega'_k (E_n)t \right) \nonumber \\ & & + {\sin \left( 
\omega'_k (E_m)t \right) \sin \left( \omega'_k (E_n)t \right) 
\over \omega'_k (E_m) \omega'_k (E_n)} (\omega^2_k + E_m E_n 
C^2_k) \Bigg] \rho^s_{nm} (0) , \label{twolv7}
\end{eqnarray}
as also obtained by Shao {\it et al}. \cite{sgc96}. We can see from Eq. (\ref{twolv7}) that the reduced density matrix of the system is independent of the temperature and squeezing conditions of the reservoir, as may be expected from the structure of the Hamiltonian
(\ref{twolv1}). This brings out the intrinsic difference between a bosonic bath of harmonic oscillators and a bath of two-level systems.

\section{Decoherence dynamics}

Here onwards we consider only the bosonic bath of harmonic 
oscillators in an initial squeezed thermal state. In this 
section, we analyze the decay of coherence of our generic 
system under various environmental conditions. We examine the 
decoherence term in the master equation for the reduced density 
matrix of the system. We also compute the dynamics of the 
`measure of coherence' related to the linear entropy of the 
system for the zero- as well as the high-temperature cases for 
different degrees of squeezing of the bath. 

\subsection{Dephasing in the system master equation}

In this subsection we examine in detail the term $\gamma(t)$ 
[Eq. (\ref{h4})]. This is the term whose time derivative is the 
decoherence-causing term as is evident from the master equation 
(\ref{h2}). To proceed, we assume a `quasi-continuous' bath 
spectrum with spectral density $I(\omega)$ such that 
\begin{equation}
\sum\limits_k {g^2_k \over \hbar^2} f(\omega_k) \longrightarrow 
\int\limits^{\infty}_0 d\omega I(\omega) f(\omega), \label{h5} 
\end{equation}
and using an Ohmic spectral density
\begin{equation}
I(\omega) = {\gamma_0 \over \pi} \omega e^{-\omega/\omega_c}, 
\label{h6} 
\end{equation}
where $\gamma_0$ and $\omega_c$ are two bath parameters, we 
obtain $\eta (t)$ in (\ref{h3}) as 
\begin{equation}
\eta (t) = -{\gamma_0 \over \pi} \tan^{-1} (\omega_c t). 
\label{h7} 
\end{equation}
In the limit $\omega_c t \gg 1$,~~ $\tan^{-1}(\omega_c t) 
\longrightarrow {\pi \over 2}$ and $\eta (t) \longrightarrow - 
{\gamma_0 \over 2}$. Now we evaluate $\gamma (t)$ given in 
(\ref{h4}) for the squeezed thermal bath for the cases of 
zero-$T$ and high-$T$. \\ \ \\ 
\underline{$T$ = 0}: \\
Using Eqs. (\ref{h5}), (\ref{h6}) in Eq. (\ref{h4}) and using 
the zero-$T$ limit we obtain $\gamma (t)$ as 
\begin{eqnarray}
\fl \gamma (t) & = & {\gamma_0 \over 2\pi} \cosh (2r) \ln 
(1+\omega^2_c t^2) - {\gamma_0 \over 4\pi} \sinh (2r) \times 
\ln \left[ {\left( 1+4\omega^2_c(t-a)^2\right) \over \left( 1+ 
\omega^2_c (t-2a)^2 \right)^2} \right]\nonumber\\ & & - 
{\gamma_0 \over 4\pi} \sinh (2r) \ln (1+4a^2\omega^2_c) , 
\label{h8} 
\end{eqnarray}
where $t> 2a$. 
Here we have taken, for simplicity, the squeezed bath 
parameters as 
\begin{eqnarray} 
\cosh \left( 2r(\omega) \right) & = & \cosh (2r),~~ \sinh 
\left( 2r (\omega) \right) = \sinh (2r), \nonumber\\ \Phi 
(\omega) & = & a\omega, \label{h13} 
\end{eqnarray} 
where $a$ is a constant depending upon the squeezed bath. The 
decoherence-causing term $\dot{\gamma} (t)$ is obtained from 
the above equation as 
\begin{equation}
\fl \frac{d \gamma (t)}{dt} = {\gamma_0 \over \pi} \cosh (2r) 
{\omega^2_c t \over (1 + \omega^2_c t^2)} - {\gamma_0 \over 
4\pi} \sinh (2r) \left[ {8\omega^2_c(t - a) \over 1 + 
4\omega^2_c (t-a)^2}-{4 \omega^2_c (t - 2a) \over 1 + 
\omega^2_c (t - 2a)^2} \right]. \label{h9} 
\end{equation}

\begin{figure}[htbp]
\begin{center}
\scalebox{1.2}{\includegraphics{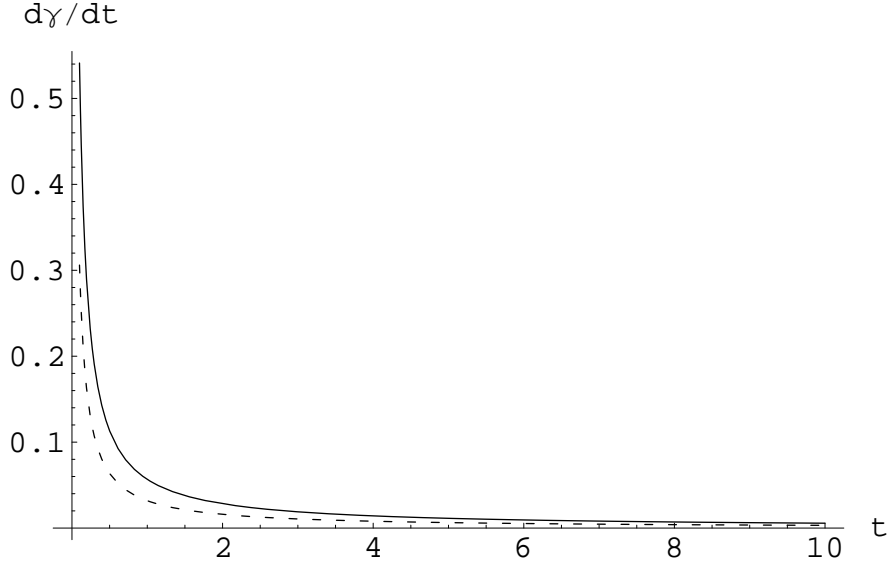}}
\end{center}
\caption{\scriptsize $\frac{d \gamma (t)}{dt}$ [Eq. (\ref{h9})] 
as a function of time $t$ for different environmental 
conditions. Here $\gamma_0 = 0.1$, $\omega_c = 50$, $a = 0$ and 
temperature $T$ = 0. The dashed and the solid curves correspond 
to the environmental squeezing parameter [Eq. (\ref{h13})] $r$ 
= 0 and 0.4, respectively. } \label{fig:gdot} 
\end{figure} 

We can see from the above equation that in the long time limit, 
$\dot{\gamma} (t) \longrightarrow \gamma_0 \cosh (2r)/(\pi t)$, 
and the terms proportional to the sine hyperbolic 
function, coming from the nonstationarity of the squeezed bath, 
are washed out. For the case of zero squeezing, we obtain from 
(\ref{h8}) 
\begin{equation}
\gamma (t) = {\gamma_0 \over 2\pi} \ln (1+\omega^2_c t^2) 
\longrightarrow {\gamma_0 \over 2\pi} \times \hbox{constant}, 
\label{h10} 
\end{equation}
because of the slow logarithmic behavior. \\
As $\omega_c \longrightarrow \infty, \gamma (t)$ in (\ref{h8}) 
tends to 
\begin{equation}
\gamma (t) \longrightarrow {\gamma_0 \over 2\pi} \cosh (2r)A - 
{\gamma_0 \over 4\pi} \sinh (2r)B , \label{h11} 
\end{equation}
where \\
$A = \lim\limits_{\omega_c \longrightarrow \infty} \ln (1 + 
\omega^2_c t^2)$ = constant, because of the slow logarithmic 
behavior and \\ $B = \lim\limits_{\omega_c \longrightarrow 
\infty} \ln \left[ {\left( 1 + 4\omega^2_c (t-a)^2 \right) 
\left( 1+4a^2\omega^2_c \right) \over \left( 1+\omega^2_c (t-
2a)^2 \right)^2} \right]$ = constant, again because of the slow 
logarithmic behavior. \\ \ \\ 
\underline{High $T$}: \\
Using (\ref{h5}), (\ref{h6}) in (\ref{h4}) and using the 
high-$T$ limit, we obtain 
\begin{eqnarray} 
\gamma (t) & = & {\gamma_0 k_BT \over \pi \hbar \omega_c} \cosh 
(2r) \left[ 2\omega_c t \tan^{-1} (\omega_c t) + \ln \left( {1 
\over 1+\omega^2_c t^2} \right) \right] \nonumber \\ & & - 
{\gamma_0 k_BT \over 2\pi \hbar \omega_c} \sinh (2r) \Bigg[ 
4\omega_c (t-a) \tan^{-1} \left( 2\omega_c (t-a) \right) 
\nonumber \\ & & - 4\omega_c (t-2a) \tan^{-1} \left( \omega_c 
(t-2a) \right) + 4a\omega_c \tan^{-1} \left( 2a\omega_c \right) 
\nonumber \\ & & + \ln \left( {\left[ 1+\omega^2_c (t-2a)^2 
\right]^2 \over \left[ 1+4\omega^2_c (t-a)^2 \right]} \right) + 
\ln \left( {1 \over 1+4a^2\omega^2_c} \right) \Bigg] , 
\label{h12} 
\end{eqnarray} 
where $t> 2a$. From (\ref{h12}) we can obtain $\gamma (t)$ for 
high $T$ and thermal bath with no squeezing, by setting $r$ and 
$a$ to zero, as 
\begin{equation} 
\gamma (t) = {\gamma_0 k_BT \over \pi \hbar \omega_c} \left[ 
2\omega_c t \tan^{-1} (\omega_c t) + \ln \left( {1 \over 
1+\omega^2_c t^2} \right) \right], \label{h14} 
\end{equation} 
such that 
\begin{equation} 
\frac{d \gamma (t)}{dt} = {2\gamma_0 k_B T \over \pi \hbar} 
\tan^{-1} (\omega_c t). \label{h15} 
\end{equation} 
This matches with the result obtained in Ref. \cite{gkd01}. For 
the case of a squeezed thermal bath, we can obtain 
$\dot{\gamma} (t)$ from (\ref{h12}) as 
\begin{eqnarray} 
\fl \frac{d \gamma (t)}{dt} & = & {2\gamma_0 k_BT \over \pi \hbar} 
\cosh (2r)\tan^{-1} (\omega_c t)-{2\gamma_0 k_BT \over \pi 
\hbar} \sinh (2r) \Bigg[ \tan^{-1} (2\omega_c (t-a)) \nonumber 
\\ & & - \tan^{-1} (\omega_c (t-2a)) \Bigg]. \label{h16} 
\end{eqnarray} 

\begin{figure}[htbp]
\begin{center}
\scalebox{1.2}{\includegraphics{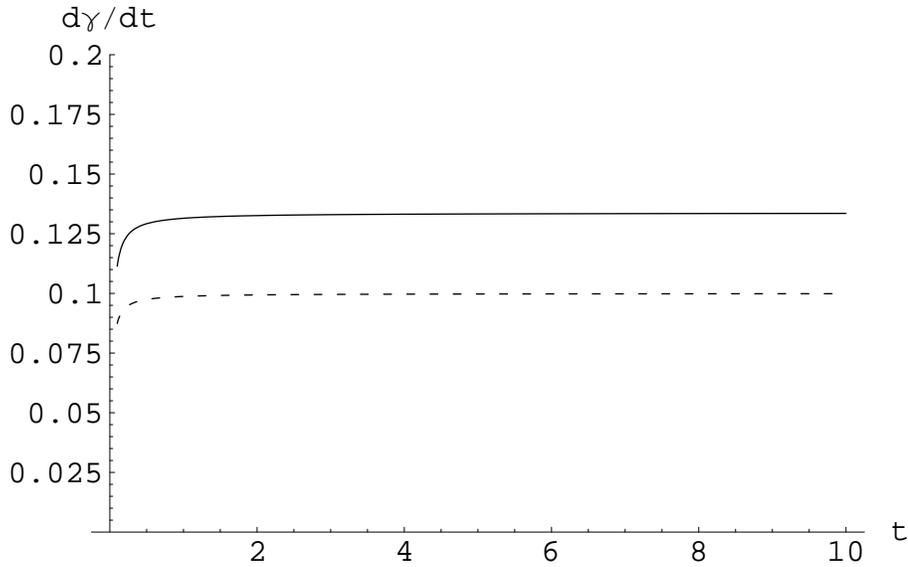}}
\end{center}
\caption{\scriptsize $\frac{d\gamma (t)}{dt}$ [Eq. (\ref{h16})] 
as a function of time $t$ for different environmental 
conditions. Here $\gamma_0 = 0.1$, $\omega_c = 50$, $a = 0$ and 
temperature $T$ (in units where $\hbar \equiv k_B \equiv 1$) = 
300. The dashed and the solid curves correspond to the 
environmental squeezing parameter [Eq. (\ref{h13})] $r$ = 0 and 
0.4, respectively.} \label{fig:gdothc} 
\end{figure} 

Figure 1 depicts the behavior of the decoherence-causing term,
$\dot{\gamma} (t)$ [Eq. (\ref{h9})], for $T = 0$ while Fig. 2 depicts
its behavior for high-$T$ [Eq. (\ref{h16})], with and without bath
squeezing indicated by the parameter $r$. A comparison between the two clearly indicates the power-law behavior of the decay of coherences at $T = 0$ and an exponential decay at high $T$. As $\omega_c \longrightarrow \infty$, from (\ref{h12}) we get
\begin{equation} 
\gamma (t) \longrightarrow {\gamma_0 k_BT \over \hbar} \cosh 
(2r) t - 2{\gamma_0 k_B T \over \hbar} \sinh (2r) a. 
\label{h17} 
\end{equation} 

\subsection{Evolution of the linear entropy -- measure of 
coherence} 

It is well-known that for a pure state $\rho^2 (t) = \rho (t)$ 
and $\rm{Tr}[\rho^2(t)] =$ 1. A mixed state instead is defined by 
the class of states which satisfies the inequality 
$\rm{Tr}[\rho^2(t)] \le$ 1. This leads to the definition of the 
linear entropy of a quantum state, $S(t) \equiv 1 - \rm{Tr} 
[\rho^2(t)]$, which is positive: $S(t) \ge$ 0, and bounded: 
$S(t) \le$ 1. $S(t)$ = 0 for a pure state and 1 for a 
completely mixed state. We can set up a related `measure of 
coherence' of the system following Ref. \cite{sgc96} as 
\begin{equation}
C(t) \equiv \rm{Tr} \left[ \rho^s (t) \right]^2. \label{h18} 
\end{equation}
If we assume the system to start from a pure state, 
\begin{equation} 
\rho^s (0) = \left[ \sum\limits_{n} p_n |n\rangle \right] 
\left[ \sum\limits_{m} p^*_m \langle m|\right], 
\end{equation} 
then using (\ref{h1}) we have 
\begin{equation} 
C(t) = \sum\limits_{m,n} |p_n|^2 |p_m|^2 e^{-2(E_n-E_m)^2 
\gamma (t)}, \label{h19} 
\end{equation} 
where $\gamma (t)$ is as in (\ref{h4}). \\ The linear entropy 
$S(t)$ can be computed from $C(t)$ as: 
\begin{equation} 
S(t) = \rm{Tr} \left[ \rho^s (t) - (\rho^s (t))^2 \right] = {\cal 
I} - C(t). \label{slin} 
\end{equation} 
$S(t)$ is plotted in Figs. 3 and 4 for a harmonic oscillator 
system starting out in a coherent state $|\alpha \rangle$ 
\cite{sz97}, for temperatures $T$ = 0 and 300, respectively, 
and for various values of environmental squeezing parameter 
$r$. \\ \ \\ 
\underline{$T$ = 0}: \\ 
Using Eq. (\ref{h4}) in Eq. (\ref{h19}) and applying the $T = 
0$ limit, i.e., making use of (\ref{h8}), the measure of 
coherence is obtained as 
\begin{eqnarray} 
C(t) & = & \sum\limits_{n,m} |p_n|^2 |p_m|^2 (1 + \omega^2_c 
t^2)^{- \gamma_0 \cosh (2r)(E_n - E_m)^2 /\pi} \nonumber \\ 
& & \times \left[ {\left( 1 + 4\omega^2_c (t - a)^2\right) \over 
\left( 1 + \omega^2_c (t - 2a)^2 \right)^2} \right]^{\gamma_0 
\sinh (2r)(E_n - E_m)^2/(2 \pi )} \nonumber \\ & & \times 
\left(1 + 4a^2 \omega^2_c \right)^{\gamma_0 \sinh (2r)(E_n - 
E_m)^2 /(2 \pi)}. \label{h20} 
\end{eqnarray} 
For the case of zero-squeezing, $r = 0 = a$, and $C(t)$ given 
by Eq. (\ref{h20}) becomes 
\begin{equation} 
C(t) = \sum\limits_{n,m} |p_n|^2 |p_m|^2 (\omega_c t)^{- 2 
\gamma_0 (E_n - E_m)^2 /\pi}. \label{h21} 
\end{equation} 

\begin{figure}[htbp]
\begin{center}
\scalebox{1.2}{\includegraphics{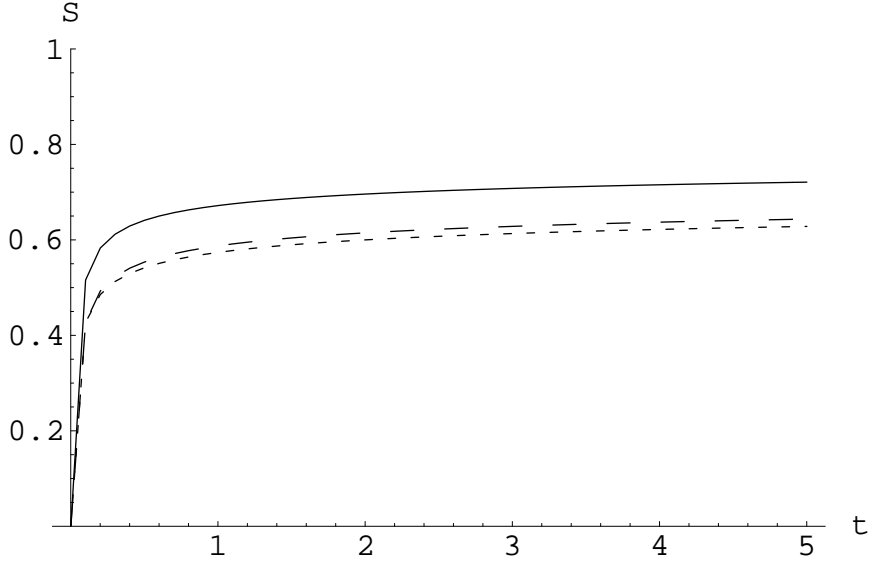}}
\end{center}
\caption{\scriptsize Linear entropy $S(t)$ [Eq. (\ref{slin})] 
as a function of time $t$ for different environmental 
conditions: $\gamma_0 = 0.1$, $\omega = 1$, $\omega_c = 50$, $a 
= 0$, $|\alpha|^2 = 5$, and $T$ = 0 so that Eq. (\ref{h8}) is 
used. The large-dashed, small-dashed and the solid curves 
correspond to the environmental squeezing parameter 
[Eq. (\ref{h13})] $r$ = 0, -0.3 and 0.4, respectively. } 
\label{fig:slinT0} 
\end{figure} 

Here we have in addition imposed the condition $\omega_c t \gg 
1$, which is a valid experimentally accessible domain of time. 
This agrees with the result obtained in \cite{sgc96}. It can be 
seen from Eqs. (\ref{h20}) and (\ref{h21}) that coherences 
follow the `power law' for $T=0$. \\ \ \\ 
\underline{High $T$}: \\
Using Eq. (\ref{h4}) in Eq. (\ref{h19}) and applying the 
high-$T$ limit, i.e., using (\ref{h12}), the measure of 
coherence is obtained as 
\begin{eqnarray}
\fl C(t) & = & \sum\limits_{m,n} |p_n|^2 |p_m|^2 \exp \Bigg\{ - 
(E_n-E_m)^2 {4\gamma_0k_BT \over \pi \hbar} \nonumber \\ & & 
\times \Bigg[ \cosh (2r) \tan^{-1} (\omega_c t) - \sinh (2r) 
\tan^{-1} \left( 2\omega_c (t-a) \right) \nonumber \\ & & + 
\sinh (2r) \tan^{-1} \left( \omega_c (t-2a) \right) \Bigg] t - 
(E_n-E_m)^2 {4a \gamma_0 k_BT \over \pi \hbar} \nonumber \\ & & 
\times \sinh (2r) \left[ \tan^{-1} \left( 2\omega_c (t-
a)\right) - 2 \tan^{-1} \left( \omega_c (t-2a) \right) -\tan^{-
1} \left( 2a\omega_c \right) \right] \Bigg\} \nonumber \\ & & 
\times (1 + \omega^2_c t^2)^{2 \gamma_0 k_B T \cosh (2r)(E_n - 
E_m)^2 /(\pi \hbar \omega_c)} \nonumber \\ & & \times \left( 
{\left[ 1 + \omega^2_c (t-2a)^2 \right]^2 \over \left[ 1 + 4 
\omega^2_c (t-a)^2 \right]}\right)^{\gamma_0 k_B T \sinh (2r)
(E_n - E_m)^2 / (\pi \hbar \omega_c)} \nonumber \\ & & \times 
(1 + 4 a^2 \omega^2_c)^{- \gamma_0 k_B T \sinh (2r)(E_n - 
E_m)^2 /(\pi \hbar \omega_c)}. \label{h23} 
\end{eqnarray}

\begin{figure}[htbp]
\begin{center}
\scalebox{1.2}{\includegraphics{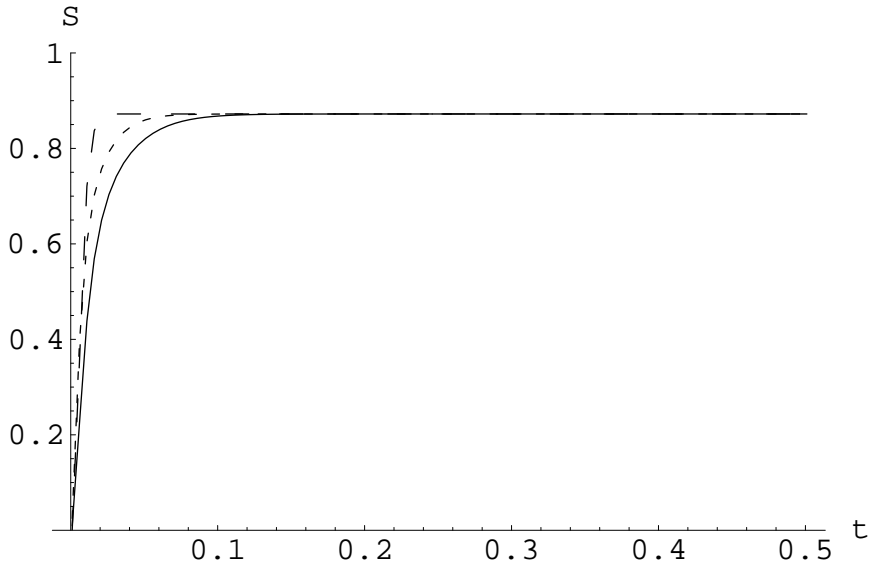}}
\end{center}
\caption{\scriptsize Linear entropy $S(t)$ [Eq. (\ref{slin})] 
as a function of time $t$ for different environmental 
conditions: $\gamma_0 = 0.1$, $\omega = 1$, $\omega_c = 50$, $a 
= 0$, $|\alpha|^2 = 5$, and $T$ (in units where $\hbar \equiv 
k_B \equiv$ 1) = 300 so that Eq. (\ref{h12}) is used. The 
solid, small-dashed and large-dashed curves correspond to the 
environmental squeezing parameter [Eq. (\ref{h13})] $r$ = 0, -
0.5 and 2, respectively. } \label{fig:slinT300} 
\end{figure} 

It can be seen from (\ref{h23}) that in the high-$T$ case, the 
measure of coherence involves exponential as well as power-law 
terms. It is also evident that the terms dominating the 
temporal behavior of the coherence measure $C(t)$ are 
\begin{eqnarray}
& \sum\limits_{n,m} & |p_n|^2 |p_m|^2 \exp \Bigg\{ - 
(E_n - E_m)^2 {4\gamma_0 k_B T \over \pi \hbar} 
\Bigg[ \cosh (2r) \tan^{-1} (\omega_c t) \nonumber \\ 
& - & \sinh (2r) \tan^{-1} \left( 2 \omega_c (t - a)\right) + 
\sinh (2r) \tan^{-1} \left( \omega_c (t - 2a) \right) \Bigg] t 
\Bigg\}. \nonumber
\end{eqnarray}
Thus in the high-$T$ limit, the behavior of the coherences is 
predominantly exponential. In the long time limit ($\omega_c t 
\rightarrow \infty$), 
\begin{equation}
C(t) \rightarrow \sum\limits_{n,m} |p_n|^2 |p_m|^2 \exp \left\{ 
- (E_n - E_m)^2 {2\gamma_0 k_BT \over \hbar} \cosh (2r) t 
\right\}. \label{h25} 
\end{equation} 
 
By comparing Figs. 3 and 4, it is evident that at $T = 0$ 
(Fig. 3), the coherences stay for a longer time characterizing 
the power-law decay as opposed to the high-$T$ case (Fig. 4), 
where the exponential decay causes the coherences to diminish 
over a much shorter period of time. Also evident is the effect 
of bath squeezing, characterized by the parameter $r$, on the 
coherences in the two temperature regimes. While in the 
zero-$T$ case, the effect of squeezing remains over a longer 
period of time, in the high-$T$ case it diminishes quickly. In 
this its behavior is similar to that of QBM of a harmonic 
oscillator system \cite{bk05, b04} at high $T$. Another 
interesting feature that comes out is that in the zero-$T$ 
regime (Fig. 3), by suitably adjusting the bath squeezing 
parameter $r$, the coherence in the system can be improved over 
the unsqueezed bath, as seen by comparing the small-dashed 
curve with the large-dashed one, representing the bath 
squeezing parameter (\ref{h13}) $r$ = -0.3 and 0, respectively. 
This clearly brings out the utility of squeezing of the thermal 
bath. 
 
\section{Comparison of the QND and non-QND evolutions and phase 
diffusion in QND} 

In this section we make a comparison between the processes 
underlying the QND and non-QND (i.e., where $[H_S, H_{SR}] \neq 
0$) types of evolution for the system of a two-level atom and a 
harmonic oscillator. We briefly consider the question of phase 
diffusion in the QND evolution of a harmonic oscillator. 

\subsection{Two-level system}

Here we take the system to be a two-level atomic system, with 
the Hamiltonian 
\begin{equation}
H_S = {\hbar \omega \over 2} \sigma_z, \label{4a1}
\end{equation}
$\sigma_z$ being the usual Pauli matrix, to be substituted in 
Eq. (1). This is a common system, with a lot of recent 
applications, as for example, in the quantum computation models 
in \cite{wu95, ps96, dd95}. \\ \ \\ 
\underline{QND evolution} \\ 
In order to study the reduced density matrix of the system under 
a QND system-reservoir interaction, i.e., for using Eq. (\ref{h1}), 
we need to identify an appropriate system eigenbasis. Here this 
is provided by the Wigner-Dicke states \cite{rd54,jr71,at72} 
$|j, m \rangle$, which are the simultaneous eigenstates of the 
angular momentum operators $J^2$ and $J_z$, and we have 
\begin{eqnarray}
H_S|j, m \rangle & = & \hbar \omega m |j, m \rangle 
\nonumber\\ & = & E_{j,m} |j, m \rangle, \label{4a2} 
\end{eqnarray} 
where $-j \le m \le j$. For the two-level system considered 
here, $j = {1 \over 2}$ and hence $m = -{1 \over 2}, {1 \over 
2}$. Using this in Eq. (\ref{h1}) and starting the system from 
the state 
\begin{equation}
|\psi(0)\rangle = \cos\left({\theta_0 \over 2}\right) |1\rangle 
+ e^{i \phi_0} \sin\left({\theta_0 \over 2}\right) |0\rangle, 
\label{4a3} 
\end{equation}
the reduced density matrix of the system after time $t$ is 
\cite{br06} 
\begin{equation}
\fl \rho^s_{m,n}(t) = 
\pmatrix {\cos^2({\theta_0 \over 2})
& {1 \over 2} \sin(\theta_0)
e^{-i (\omega t + \phi_0)} e^{-(\hbar \omega)^2 \gamma(t)}
\cr {1 \over 2} \sin(\theta_0) 
e^{i(\omega t + \phi_0)} e^{-(\hbar \omega)^2 \gamma(t)}
& \sin^2({\theta_0 \over 2})}, \label{4a4}
\end{equation}
from which the Bloch vectors can be extracted to yield
\begin{eqnarray}
\langle \sigma_x (t) \rangle &=& \sin(\theta_0) \cos(\omega t + 
\phi_0) e^{-(\hbar \omega)^2 \gamma(t)}, \nonumber\\ \langle 
\sigma_y (t) \rangle &=& \sin(\theta_0) \sin(\omega t + \phi_0) 
e^{-(\hbar \omega)^2 \gamma(t)}, \nonumber\\ \langle \sigma_z 
(t) \rangle &=& \cos(\theta_0). \label{4a5} 
\end{eqnarray}
Here $\gamma(t)$ is as in Eqs. (\ref{h8}), (\ref{h12}) for zero 
and high $T$, respectively and $\sigma_x$, $\sigma_y$, 
$\sigma_z$ are the standard Pauli matrices. It can be easily 
seen from the above Bloch vector equations that the QND 
evolution causes a coplanar, fixed by the polar angle 
$\theta_0$, in-spiral towards the $z$-axis of the Bloch sphere. 
This is the characteristic of a phase-damping channel 
\cite{nc00}. \\ \ \\ 
\underline{Non-QND evolution of the Lindblad form} \\ 
Next we study the reduced dynamics of the system (\ref{4a1}) 
interacting with a squeezed thermal bath under a weak 
Born-Markov and rotating wave approximation. This implies that 
here the system interacts with its environment via a non-QND 
interaction such that along with a loss in phase information, 
energy dissipation also takes place. The evolution has a 
Lindblad form which in the interaction picture is given by 
\cite{sz97,bp02} 
\begin{eqnarray}
{d \over dt}\rho^s(t) & = & \gamma_0 (N + 1) \left(\sigma_- 
\rho^s(t) \sigma_+ - {1 \over 2}\sigma_+ \sigma_- \rho^s(t) -{1 
\over 2} \rho^s(t) \sigma_+ \sigma_- \right) \nonumber \\ & & + 
\gamma_0 N \left( \sigma_+ \rho^s(t) \sigma_- - {1 \over 
2}\sigma_- \sigma_+ \rho^s(t) -{1 \over 2} \rho^s(t) \sigma_- 
\sigma_+ \right) \nonumber \\ & & - \gamma_0 M \sigma_+ 
\rho^s(t) \sigma_+ -\gamma_0 M^* \sigma_- \rho^s(t) \sigma_- . 
\label{4a6} 
\end{eqnarray}
Here
\begin{equation}
N = N_{th}(\cosh^2 r + \sinh^2 r) + \sinh^2 r, \label{4a7} 
\end{equation}
\begin{equation}
M = -{1 \over 2} \sinh(2r) e^{i\Phi} (2 N_{th} + 1), 
\label{4a8} 
\end{equation}
and
\begin{equation}
N_{th} = {1 \over e^{\hbar \omega /(k_B T)} - 1}, 
\label{4a9} 
\end{equation}
where $N_{th}$ is the Planck distribution giving the number of thermal
photons  at the  frequency  $\omega$, and  $r$,  $\Phi$ are  squeezing
parameters of the  bath. The case of a  thermal bath without squeezing
can be obtained from the  above expressions by setting these squeezing
parameters to  zero. $\gamma_0$ is  a constant typically  denoting the
system-environment coupling  strength, and $\sigma_+$,  $\sigma_-$ are
the standard raising and lowering operators, respectively, given by
\begin{eqnarray}
\sigma_+ & = & |1 \rangle \langle 0| = {1 \over 2} 
\left(\sigma_x + i \sigma_y \right), \nonumber \\ \sigma_- & = 
& |0 \rangle \langle 1| = {1 \over 2} \left( \sigma_x - i 
\sigma_y \right) . \label{4a10} 
\end{eqnarray}
In the above equation $|1 \rangle$ is the upper state of the 
atom and $|0 \rangle$ is the lower state. Evolving the system 
given by $H_S$ from the initial state given in Eq. (\ref{4a3}), 
using Eq. (\ref{4a6}), we obtain the reduced density matrix of 
the system from which the Bloch vectors can be extracted to 
yield \cite{br06} 
\begin{eqnarray}
\langle \sigma_x (t) \rangle & = & \left[1 + {1 \over 2} 
\left(e^{\gamma_0 a t} - 1\right) (1 + \cos \Phi)\right] e^{-
\gamma_0 (2N + 1 + a)t/2} \langle \sigma_x (0) \rangle 
\nonumber \\ & & - \sin \Phi \sinh \left( {\gamma_0 a t \over 
2} \right) e^{- \gamma_0(2N + 1)t/2} \langle \sigma_y 
(0) \rangle, \nonumber \\ \langle \sigma_y (t) \rangle & = & 
\left[1 + {1 \over 2} \left(e^{\gamma_0 a t} - 1\right) (1 - 
\cos \Phi)\right] e^{- \gamma_0 (2N + 1 + a)t/2} \langle 
\sigma_y (0) \rangle \nonumber \\ & & - \sin \Phi \sinh \left( 
{\gamma_0 a t \over 2}\right) e^{- \gamma_0 (2N + 1)t/2} 
\langle \sigma_x (0) \rangle, \nonumber \\ \langle \sigma_z (t) 
\rangle & = & e^{-\gamma_0 (2N + 1)t} \langle \sigma_z (0) 
\rangle - {1 \over (2N + 1)} \left(1 - e^{-\gamma_0 (2N + 1)t} 
\right), \label{4a11} 
\end{eqnarray}
where
\begin{equation}
a = \sinh(2r) (2 N_{th} + 1). \label{4a12}
\end{equation}
It can be seen from Eq. (\ref{4a11}) that the reduced density 
matrix $\rho^s (t)$ shrinks towards the asymptotic equilibrium 
state $\rho_{asymp}$, given by 
\begin{equation}
\rho_{asymp} = \pmatrix{1-p & 0 \cr 0 & p}, \label{4a13}
\end{equation}
where $p = \frac{1}{2}\left[1 + \frac{1}{(2N+1)}\right]$. For the 
case of zero squeezing and zero temperature, this action 
corresponds to an amplitude-damping channel \cite{nc00,br06} 
with the Bloch sphere shrinking to a point representing the 
state $|0 \rangle$ (the south pole of the Bloch sphere) while 
for the case of finite $T$ but zero squeezing, the above action 
corresponds to a generalized amplitude-damping channel 
\cite{nc00,br06} with the Bloch sphere shrinking to a point 
along the line joining the south pole to the center of the 
Bloch sphere. The center of the Bloch sphere is reached in the 
limit of infinite temperature. 

The above analysis brings out the point that while the case of the QND
system-environment interaction corresponds to a phase-damping channel,
the  case where  the evolution  is  non-QND, in  particular where  the
evolution  is generated by  Eq. (\ref{4a6}),  having a  Lindblad form,
corresponds  to a  (generalized) amplitude-damping  channel  (for zero
bath  squeezing). This  brings out  in a  very transparent  manner the
difference  in the  quantum statistical  mechanics underlying  the two
processes.  While in  the case  of QND  interaction, the  system tends
(along  the $z$-axis)  towards  a  localized state,  for  the case  of
non-QND  interaction, the  system  tends towards  a unique  asymptotic
equilibrium state, which would be pure  (for $T = 0$) or mixed (for $T
> 0$).  This  can be  seen  from  Figs. 5,  where  the  effect of  the
environment  on  the  initial  Bloch  sphere [Fig.  5(A)]  is  brought
out. Figure 5(B) depicts  the evolution under a QND system-environment
interaction [Eqs.  (\ref{4a5})] while Figs.  5(C) and 5(D)  depict the
evolution   under   a   dissipative   system-environment   interaction
[Eqs.  (\ref{4a11})]. While  Fig.  5(B) clearly  shows  a tendency  of
localization along  the $z$-axis, Figs.  5(C) and 5(D)  illustrate the
tendency  of  going  towards  a  unique  asymptotic  fixed  point.  In
Fig. 5(D), the  presence of a finite $\Phi$  (\ref{4a8}) is manifested
in the tilt in the figure.

\begin{figure}
\includegraphics[width=6.5cm]{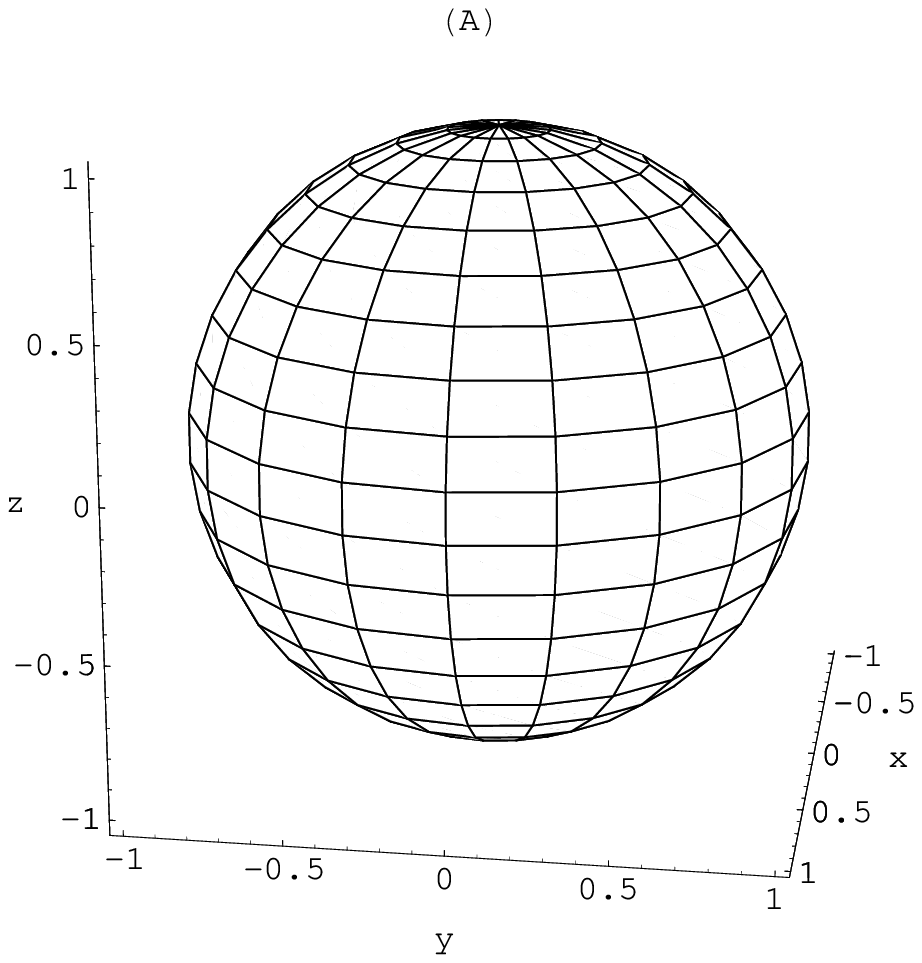}\hspace*{1.0cm}
\includegraphics[width=5.5cm]{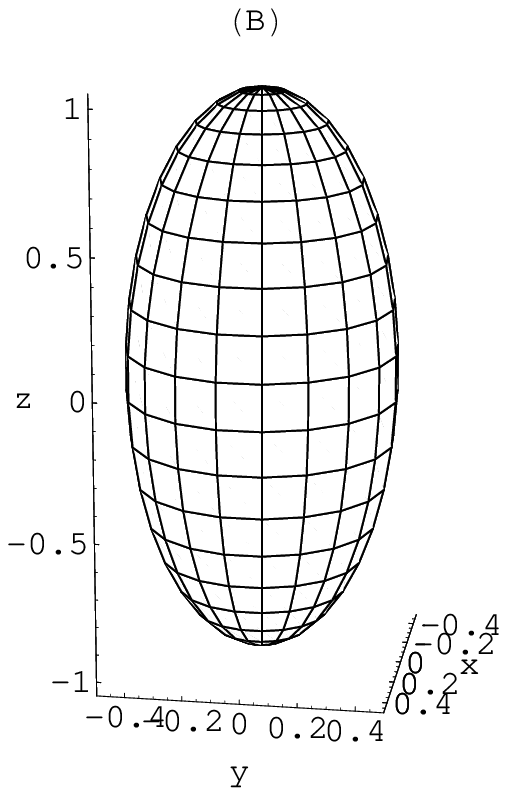} \\
\includegraphics[width=7.0cm]{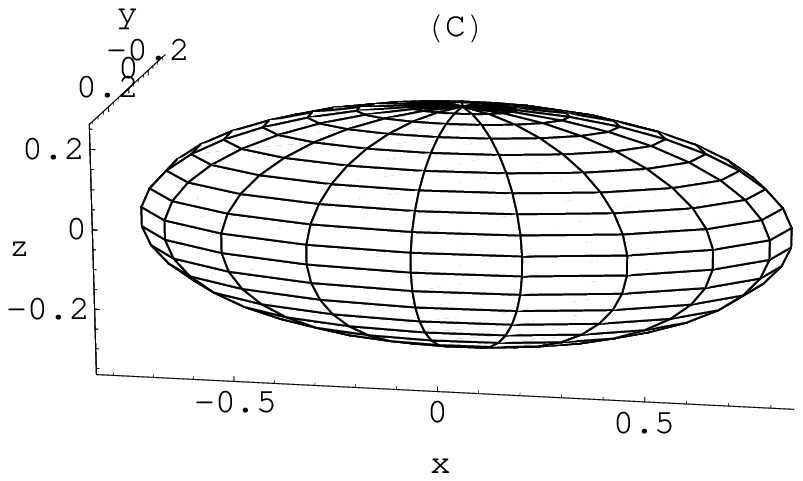}\hspace*{1.0cm}
\includegraphics[width=7.0cm]{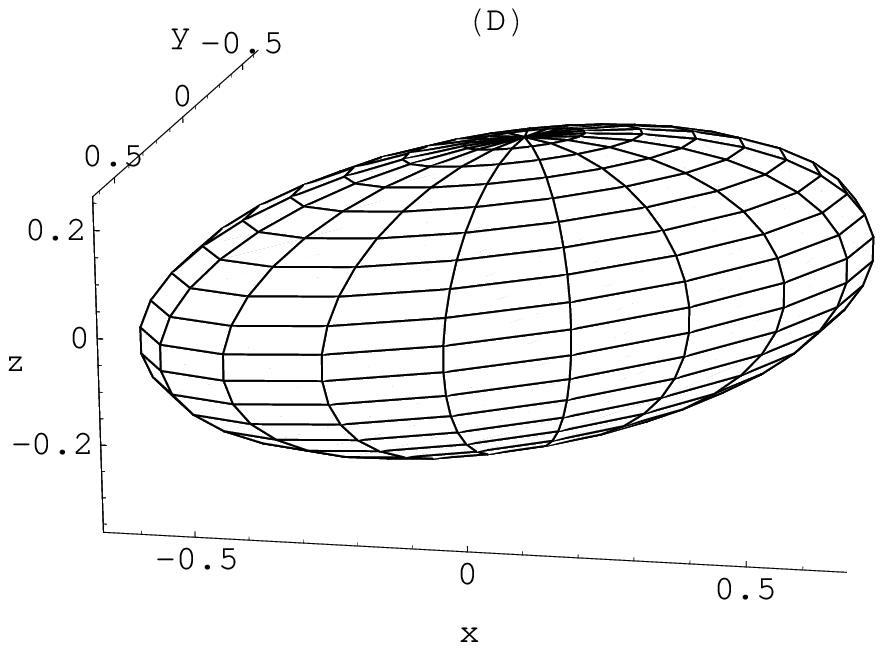}
\caption{\scriptsize Effect of QND and dissipative interactions 
on the Bloch sphere: (A) the full Bloch sphere; (B) the Bloch 
sphere after time $t = 20$, with $\gamma_0 = 0.2$, $T = 0$, 
$\omega = 1$, $\omega_c = 40\omega$ and the environmental 
squeezing parameter [Eq. (\ref{h13})] $r = a = 0.5$, evolved 
under a QND interaction [Eqs. (\ref{4a5})]; (C) and (D) the 
effect of the Born-Markov type of dissipative interaction 
[Eqs. (\ref{4a11})] with $\gamma_0 = 0.6$ and temperature $T = 
5$, on the Bloch sphere -- the $x$ and $y$ axes are 
interchanged to present the effect of squeezing more clearly. 
(C) corresponds to $r = 0.4$, $\Phi = 0$ and $t = 0.15$ while 
(D) corresponds to $r = 0.4$, $\Phi = 1.5$ and $t = 0.15$. } 
\end{figure} 

\subsection{Harmonic oscillator system}

Next we take a system of harmonic oscillator with the 
Hamiltonian 
\begin{equation} 
H_S = \hbar \omega \left(a^{\dag}a + {1 \over 2} \right).
\label{h26}
\end{equation}
The Hamiltonian $H_S$ (\ref{h26}), substituted in Eq. (1), has 
been used by Turchette {\it et al.} \cite{tw00} to describe an 
experimental study of the decoherence and decay of quantum 
states of a trapped atomic ion's harmonic motion interacting 
with an engineered `phase reservoir'. \\ \ \\ 
\underline{QND evolution} \\ Noting that the master equation 
(\ref{h2}) in the system space can be written equivalently as 
\begin{equation}
\dot{\rho}^s = -{i \over \hbar} [H_S, \rho^s] + i \dot{\eta}(t) 
[H^2_S, \rho^s] - \dot{\gamma}(t)\left(H^2_S \rho^s - 2 H_S 
\rho^s H_S + \rho^s H^2_S \right), \label{h27} 
\end{equation}
and substituting (\ref{h26}) in (\ref{h27}) we obtain the 
master equation for a harmonic oscillator coupled to a bosonic 
bath of harmonic oscillators by a QND type of coupling as 
\begin{eqnarray}
\dot{\rho}^s & = & -i \omega [a^{\dag}a, \rho^s] + i \hbar^2 
\omega^2 \dot{\eta}(t) \left[(a^{\dag}a)^2 + a^{\dag}a, \rho^s 
\right] \nonumber \\ & & - \hbar^2 \omega^2 \dot{\gamma}(t)
\left[(a^{\dag}a)^2 \rho^s - 2 a^{\dag}a \rho^s a^{\dag}a + 
\rho^s (a^{\dag}a)^2 \right]. \label{h28} 
\end{eqnarray} 
For clarity we transform the above equation into the form of a 
$Q$ distribution function \cite{sz97} given by the prescription 
\begin{equation}
Q(\alpha) = {1 \over \pi} \langle \alpha|\rho| \alpha \rangle, 
\label{h29} 
\end{equation} 
where $|\alpha\rangle$ is a coherent state. From the master 
equation (\ref{h28}), the equation for the $Q$ distribution 
function becomes 
\begin{eqnarray}
\fl {\partial  \over \partial  t} Q  & = &  -i \omega  \left( \alpha^*
{\partial \over \partial \alpha^*  } - \alpha {\partial \over \partial
\alpha }  \right) Q  + i \hbar^2  \omega^2 \dot{\eta}(t) \Bigg[  2(1 +
\alpha \alpha^*) \left( \alpha^* {\partial \over \partial \alpha^* } -
\alpha {\partial  \over \partial \alpha }  \right) \nonumber \\  & & +
\left(  \alpha^{*  2}  {\partial^2  \over \partial  \alpha^{*  ^2}}  -
\alpha^2  {\partial^2  \over \partial  \alpha^2}  \right)  \Bigg] Q  -
\hbar^2  \omega^2  \dot{\gamma}(t)  \nonumber  \\ &  &  \times  \left[
\alpha^* {\partial  \over \partial \alpha^* }+  \alpha {\partial \over
\partial \alpha } +  \alpha^{* 2} {\partial^2 \over \partial \alpha^{*
2}}  +  \alpha^2  {\partial^2  \over  \partial \alpha^2}  -  2  \alpha
\alpha^* {\partial^2 \over  \partial \alpha \partial \alpha^*} \right]
Q. \label{h30}
\end{eqnarray}
Using polar coordinates $\alpha = \xi e^{i\theta}$, this 
equation can be transformed into 
\begin{eqnarray}
{\partial \over \partial t}Q &=& \omega {\partial \over 
\partial \theta}Q - \hbar^2 \omega^2 \dot{\eta}(t) \left[ (1+ 2 
{\xi}^2){\partial \over \partial \theta} + \xi {\partial^2 
\over \partial{\xi} \partial{\theta}} \right]Q \nonumber \\ & & 
+ \hbar^2 \omega^2 \dot{\gamma}(t) {\partial^2 \over 
\partial{\theta}^2}Q. \label{h31} 
\end{eqnarray} 

\noindent \underline{Non-QND  QBM} \\ To compare  (\ref{h30}) with the
equation  obtained  in  the  case  of  QBM of  a  system  of  harmonic
oscillator interacting with a squeezed thermal bath, we start with the
general QBM master equation \cite{bk05,b04}:
\begin{eqnarray}
\fl i\hbar {\partial \over \partial {t}} \rho^{s}(x,x',t) & = & 
\left\{ \frac{-\hbar^2}{2M} \left({\partial^2 \over 
\partial{x}^2 }-{\partial^2 \over \partial{x'}^2}\right) 
+\frac{M}{2} \Omega_{ren}^2 (t) (x^2 - x'^2)\right\} 
\rho^{s}(x,x',t) \nonumber \\ & & -i\hbar \Gamma(t)(x - x') 
\left( {\partial \over \partial{x}} - {\partial \over 
\partial{x'}}\right) \rho^{s}(x,x',t) \nonumber \\ & & + i 
D_{pp}(t) (x - x')^2 \rho^{s}(x,x',t) \nonumber \\ & & - 
\hbar(D_{xp}(t) + D_{px}(t))(x - x') \left({\partial \over 
\partial{x}} + {\partial \over \partial{x'}}\right) 
\rho^{s}(x,x',t) \nonumber \\ & & - i \hbar^2 D_{xx}(t) 
\left({\partial \over \partial{x}} + {\partial \over 
\partial{x'}}\right)^2 \rho^{s}(x,x',t) . \label{h32} 
\end{eqnarray}
Here $\Gamma(t)$ is the term responsible for dissipation, 
$D_{pp}(t)$ for decoherence, $D_{xx}(t)$ promotes diffusion in 
$p^2$, and $D_{xp}(t)$, $D_{px}(t)$ are responsible for 
promoting (anomalous) diffusion in $xp + px$. The details of 
these coefficients of the master equation (\ref{h32}) can be 
found in \cite{bk05,b04}. Here the coordinate representation 
of the density matrix has been used in contrast to the energy 
representation used in (\ref{h2}). Comparing (\ref{h32}) with 
(\ref{h2}) we find that the QND coupling of the system with the 
environment makes the quantum statistical mechanics of the 
evolution much simpler. As already noted below Eq. (\ref{h4}), 
a comparison between (\ref{h2}) and (\ref{h32}) shows that in 
the QND case there is a decoherence-governing term 
$\dot{\gamma}(t)$, but no term responsible for dissipation. In 
contrast, the QBM case has dissipation and a number of {\it 
diffusion channels} as seen by the existence of the diffusion 
terms $D_{xx}(t)$, $D_{xp}(t)+ D_{px}(t)$ and $D_{pp}(t)$. 

Since Eq. (\ref{h32}) is also obtained for a harmonic 
oscillator system (cf. (\ref{h26})), we proceed as before and 
obtain its corresponding $Q$ equation as 
\begin{eqnarray}
i\hbar{\partial \over \partial t}Q & = & \hbar \omega \left[ 
\alpha^{*} {\partial \over \partial \alpha^{*} } - \alpha 
{\partial \over \partial \alpha } \right] Q \nonumber \\ & & + 
{i \hbar \over 2} \Gamma(t) \Bigg[ - {\partial^2 \over \partial 
\alpha^{* 2} } - {\partial^2 \over \partial \alpha^2} + 2 
{\partial^2 \over \partial \alpha \partial \alpha^{*}} 
\nonumber \\ & & - 2 \alpha {\partial \over \partial 
\alpha^{*}} - 2 \alpha^{*} {\partial \over \partial \alpha } + 
2 \alpha^{*}{\partial \over \partial \alpha^{*}} + 2 \alpha 
{\partial \over \partial \alpha} + 4 \Bigg] Q \nonumber \\ & & 
+ {i \hbar \over m \omega} D_{pp}(t) \left[- { \partial^2 \over 
\partial \alpha \partial \alpha^{*}}+ {1 \over 2} {\partial^2 
\over \partial \alpha^{* 2} } + {1 \over 2} {\partial^2 \over 
\partial \alpha^2} \right] Q \nonumber \\ & & - {\hbar \over 2} 
(D_{xp}(t) + D_{px}(t)) \left[ {\partial^2 \over \partial 
\alpha^{* 2} } - {\partial^2 \over \partial \alpha^2} \right] Q 
\nonumber \\ & & - i\hbar m \omega D_{xx}(t) \left[ {\partial^2 
\over \partial \alpha \partial \alpha^{*}} + {1 \over 2} 
{\partial^2 \over \partial \alpha^{* 2} } + {1 \over 2} 
{\partial^2 \over \partial \alpha^2} \right] Q. \label{h33} 
\end{eqnarray}
From  a comparison  of Eq.  (\ref{h33})  with Eq.  (\ref{h30}), it  is
evident that  QBM is  a more complicated  process than  QND evolution.
Writing  Eq. (\ref{h33}) in  polar coordinates  does not  simplify its
structure, unlike the case of  QND evolution where Eq. (\ref{h31}) was
obtained in a  simple form in polar coordinates.  This is a reflection
of the fact that QBM is a more complicated process than QND as well as
the  fact that  in the  QND case,  the master  equation  (\ref{h2}) is
obtained  in  the  system  energy  basis which  is  more  amenable  to
simplification  in  the  $Q$  representation (the  $Q$-function  being
proportional  to the  diagonal element  of the  density matrix  in the
coherent state basis) than  the coordinate representation in which the
QBM master equation (\ref{h32})  is obtained. \\ \ \\ \underline{Phase
diffusion  of   the  QND  harmonic  oscillator}  \\   We  now  analyze
Eq.  (\ref{h31})  to gain  some  insight  into  the process  of  phase
diffusion in the  case of a harmonic oscillator  system coupled to its
bath via a QND type of coupling.  We take the long time limit. In this
limit   $\dot{\eta}(t)   \longrightarrow    0$   (cf.   remark   below
(\ref{h7})). We solve for the Q distribution function for the zero and
high temperature  cases. \\ \ \\  \underline{$T$ = 0}: \\  In the long
time  limit, $\dot{\eta}(t)  \longrightarrow 0$,  and $\dot{\gamma}(t)
\longrightarrow 0$ (cf. Fig. 1). Then Eq. (\ref{h31}) reduces to
\begin{equation} 
{\partial \over \partial t}Q = \omega {\partial \over \partial 
\theta}Q, \label{h35} 
\end{equation}
which has the solution
\begin{equation}
Q(\theta, t) = e^{-\lambda t}e^{- \lambda \theta / \omega}, 
\label{h36} 
\end{equation}
where $\lambda$ is a constant. Equation (\ref{h35}) does not have 
the form of a standard diffusion equation in phase space -- there 
is a drift term but no diffusion term. \\ \ \\ 
\underline{High $T$}: \\
In the long time limit, $\dot{\eta}(t) \longrightarrow 0$ and 
$\dot{\gamma}(t) \longrightarrow \gamma_0 k_B T \cosh(2r)/ \hbar$ 
(as can be inferred from Eq. (\ref{h16}) and 
Fig. 2). Then Eq. (\ref{h31}) becomes 
\begin{equation} 
{\partial \over \partial t}Q = \omega {\partial \over \partial 
\theta}Q + A_1 {\partial^2 \over \partial{\theta}^2}Q ,
\label{h37} 
\end{equation}
with
\[ A_1 = \hbar {\omega}^2 \gamma_0 k_B T \cosh(2r). \]
This has the elementary solution 
\begin{equation}
Q(\theta, t)= e^{-\alpha t} e^{-A \theta}\left[c_1 e^{B \theta} 
+ c_2 e^{-B \theta} \right], \label{h38} 
\end{equation}
where $\alpha$, $c_1$, $c_2$ are constants, 
\[A = {\omega \over 2 A_1} , \]
and 
\begin{equation}
B = {\omega \over 2 A_1} {\sqrt{1- {4\alpha A_1 \over 
\omega^2}}}. \label{h39} 
\end{equation}
Equation (\ref{h37}) has  the form of a time-dependent  diffusion on a
circle. It does  not have the form of a pure  diffusion because of the
presence  of  an additional  Kerr-like  term  in  the master  equation
(\ref{h28}).  A  form similar  to this arises  in the  phase diffusion
model for the phase fluctuations of  the laser field when the laser is
operated far above threshold so that the amplitude fluctuations can be
ignored  \cite{sz97}.   Then  the  phase fluctuations  due  to  random
spontaneous emissions can be modelled as a one-dimensional random walk
along the angular direction. In this sense it can be said that the QND
Hamiltonian describes  diffusion of  the quantum phase  \cite{pb89} of
the light field. From Eq. (\ref{h37}) it is evident that the diffusion
coefficient  is  dependent  on   temperature  $T$  and  the  reservoir
squeezing parameter $r$. In the  high temperature and long time limits
the  dynamical  behavior  is  that  of  a  quantum  mechanical  system
influenced  by  an  environment   that  is  modelled  by  a  classical
stochastic process, a situation that was studied in \cite{gkd01}. That
this is  not so for  the zero-$T$ case  suggests that a  zero-$T$ open
quantum system cannot be simulated, even  in the long time limit, by a
classical stochastic bath. A  detailed analysis of the phase diffusion
pattern in  QND types of evolution  for the two-level atom  as well as
the harmonic oscillator system has been given in Ref. \cite{sb07-2}.

\section{Conclusions}

In this paper  we have studied the dynamics of  a generic system under
the influence of  its environment where the coupling  of the system to
its  environment is  of the  energy-preserving QND  type. The  bath is
initially in a  squeezed thermal state, decoupled from  the system. We
have compared the QND results with the case where the coupling is of a
non-QND dissipative type for a system of two-level atom and a harmonic
oscillator.

For  a  bosonic bath  of  harmonic  oscillators  with a  QND  coupling
(section  2.1), we  have  found that  in  the master  equation of  the
system,  though there  is a  term governing  decoherence, there  is no
dissipation  term,  i.e.,  such  systems undergo  decoherence  without
dissipation of energy. For the case where there is no squeezing in the
bath,     our    results     reduce    to     those     obtained    in
Refs.  \cite{sgc96,mp98,gkd01} for the  case of  a thermal  bath.  The
reduced  density matrix  of  the  system interacting  with  a bath  of
two-level systems (section 2.2) via a QND type of coupling is found to
be independent of temperature \cite{sgc96} and squeezing conditions of
the bath.  This brings out  an intrinsic difference between  a bosonic
bath of harmonic oscillators and a bath of two-level systems.

We have  analyzed the effect of  the phase-sensitivity of  the bath on
the  dynamics of  decoherence, first  by looking  at the  term causing
decoherence  in the  system master  equation for  the bosonic  bath of
harmonic oscillators (obtained in  section 2.1). We have evaluated the
decoherence-causing term for the  cases of zero and high temperatures,
and also obtained  its long time limit for both the  cases. A study of
the linear entropy $S(t)$, which  is an indicator of the coherences in
the  reduced density matrix  of the  system, clearly  reveals (section
3.2) that in  the high-$T$  case, the effect  of the squeezing  in the
bath is quickly washed out and  the system loses coherence over a very
short time  scale. In contrast,  in the zero-$T$ case,  the coherences
are preserved  over a longer period  of time and the  squeezing in the
bath can actually  be used to improve the  coherence properties of the
system.
 
We have  made a comparison between the  quantum statistical mechanical
processes   of   QND   and   non-QND   types   of   system-environment
interaction. For a  two-level atomic system (section 4.1),  it is seen
that  whereas the  action  of the  QND system-environment  interaction
tends to localize the system  along the $z$-axis indicative of, in the
parlance  of  quantum  information  theory,  a  phase-damping  channel
\cite{nc00},  the  non-QND  interaction  (epitomized by  the  Lindblad
equation  (\ref{4a6}))  tends to  take  the  system  towards a  unique
asymptotic  fixed point,  which for  the case  of zero  bath squeezing
would  be indicative  of the  (generalized)  amplitude-damping channel
\cite{nc00}. For  a harmonic oscillator system (section  4.2), we have
converted   the  master  equation   to  the   equation  for   the  $Q$
representation.  This  brings  about  in  a very  general  manner  the
differences in  the quantum statistical  mechanical processes involved
in QND and in QBM. The QBM  process is much more involved than the QND
one in that  in addition to the decoherence  and dissipation terms, it
contains a number of other diffusion terms. In our analysis of the QND
equation for the  harmonic oscillator system, in the  long time limit,
we find a form similar to the one in the phase diffusion model for the
fluctuations  of  the laser  operated  far  above  threshold when  the
amplitude fluctuations can be  ignored.  The phase fluctuations due to
random spontaneous emissions can be modeled as a random walk along the
angular  direction.  In  this  sense  the  QND  Hamiltonian  describes
diffusion of the quantum phase of  the light field. We find that while
in  the  high-$T$ case  the  situation can  be  modeled  as a  quantum
mechanical system influenced by  a classical stochastic process, it is
not so for the zero-$T$  case. The high-$T$ $Q$ equation resembles the
equation  of  phase  diffusion  on  a circle  which  would  suggest  a
connection   between  quantum  phase   diffusion  and   QND  evolution
\cite{sb07-2}.  Our   quantitative  study  provides   a  step  towards
understanding  and control of  the environmental  impact in  such open
quantum systems.

\ack

The  School  of Physical  Sciences,  Jawaharlal  Nehru University,  is
supported  by  the  University   Grants  Commission,  India,  under  a
Departmental Research Support scheme.

\appendix

\section{Derivation of reduced density matrix (Eq. (\ref{h1}))
\label{sec:deriv}}

In this  appendix, we present the  steps leading to  the derivation of
the   reduced   density  matrix   $\rho^s_{nm}   (t)$   from  (4)   to
(\ref{h1}). It follows from Eq. (\ref{hn}) that
\begin{eqnarray}
e^{i H^{(k)}_n t/\hbar  } & = & e^{-i E^2_n g^2_k  \left( \omega_k t -
\sin (\omega_k t)\right)/(\hbar^2 \omega^2_k)} \nonumber \\ & & \times
D \left(  {E_n g_k  \over \hbar \omega_k}  \left( e^{i\omega_k t}  - 1
\right) \right) e^{i\omega_k b^{\dagger}_k b_k t} , \label{appn1}
\end{eqnarray}
where $D(\alpha)$ is the displacement operator,
\begin{equation}
D(\alpha) = e^{\alpha b^{\dagger}_k - \alpha^* b_k} ,
\end{equation}
and $\sum\limits_k H^{(k)}_n = H_n$. Similarly,
\begin{eqnarray}
e^{- i H^{(k)}_n t/\hbar } & =  & e^{i E^2_n g^2_k \left( \omega_k t -
\sin  (\omega_k t)  \right)/(\hbar^2  \omega^2_k)} D  \left( {E_n  g_k
\over  \hbar \omega_k}  \left( e^{-  i\omega_kt} -  1  \right) \right)
\nonumber   \\   &  &   \times   e^{-   i\omega_k  b^{\dagger}_k   b_k
t}. \label{appn2}
\end{eqnarray}
Now using  Eqs. (\ref{appn1})  and (\ref{appn2}) in  Eq. (\ref{rhot}),
and  making  use  of  the  following properties  of  the  displacement
operator
\begin{equation}
e^{i\omega_k b^{\dagger}_k b_k t} D(\alpha) = D(\alpha 
e^{i\omega_k t})~ e^{i\omega_k b^{\dagger}_k b_k t}, 
\end{equation}
\begin{equation}
D^{\dagger}(\alpha) = D(-\alpha),
\end{equation}
\begin{equation}
D^{\dagger}(\alpha)  D(\alpha   e^{i\omega_k  t})  =   D\left(  \alpha
(e^{i\omega_k t}-1) \right) e^{i\alpha \alpha^* \sin (\omega_k t)},
\end{equation}
the reduced density matrix in the system eigenbasis becomes 
\begin{eqnarray}
\rho^s_{nm}(t) &  = &  e^{- i  (E_n - E_m)t/  \hbar}~~ e^{-  i(E^2_n -
E^2_m)  \sum\limits_k (g^2_k  \sin  (\omega_k t)/\hbar^2  \omega^2_k)}
\nonumber \\ & & \times \prod_k \rm{Tr}_R \left[ \rho_R(0) D(\theta_k)
\right] \rho^s_{nm}(0). \label{appn3}
\end{eqnarray}
Here
\begin{equation}
\theta_k = (E_m - E_n) {g_k \over \hbar \omega_k} 
(e^{i\omega_kt} - 1) , 
\end{equation}
and $\rho_R(0)$ is as in Eq. (\ref{rhorin}).

The trace term in Eq. (\ref{appn3}) is
\begin{eqnarray}
\fl  \prod_k \rm{Tr}_R  \left[  \rho_R(0) D(\theta_k)  \right]  & =  &
\prod_k  \rm{Tr}_R \left[ S(r_k,  \Phi_k) \rho_{th}  S^{\dagger} (r_k,
\Phi_k)  D(\theta_k) \right]  \nonumber  \\ &  =  & \prod_k  \rm{Tr}_R
\left[  \rho_{th} D  \left( \theta_k  \cosh (r_k)  +  \theta^*_k \sinh
(r_k) e^{2i \Phi_k}  \right) \right] \nonumber \\ & =  & \exp \Bigg[ -
{1  \over  2}  (E_m   -  E_n)^2  \sum\limits_k  {g^2_k  \over  \hbar^2
\omega^2_k}  \coth  \left(  {\beta  \hbar \omega_k  \over  2}  \right)
\nonumber  \\ & &  \times \left|(e^{i\omega_k  t} -  1) \cosh  (r_k) +
(e^{-i  \omega_k   t}  -  1)  \sinh  (r_k)   e^{2i  \Phi_k}  \right|^2
\Bigg]. \label{appn4}
\end{eqnarray}
Here we have used the following relation beween squeezing and 
displacement operators: 
\begin{equation}
S^{\dagger}  (r_k, \Phi_k)  D  (\theta_k) S(r_k,  \Phi_k)  = D  \left(
\theta_k \cosh (r_k) + \theta^*_k \sinh (r_k) e^{2i\Phi_k} \right).
\end{equation}
Using  Eq. (\ref{appn4})  in  Eq. (\ref{appn3}),  the reduced  density
matrix $\rho^s_{nm} (t)$ (\ref{h1}) is obtained.

\end{document}